\documentclass[aps,prd,twocolumn,nofootinbib,amsmath,amssymb]{revtex4-2}

\usepackage{graphicx}
\usepackage{bm}
\usepackage{hyperref}

\begin{document}

\title{A General-Relativistic Effective Mapping for Path-Dependent Redshift in the Low-Redshift Universe}

\author{Said Laaroua}
\affiliation{Department of Physics, Santa Rosa Junior College, Santa Rosa, CA 95401, USA}

\begin{abstract}
We develop an effective mapping for low-redshift photon propagation that captures the leading path-dependent deviations from the standard FLRW redshift. Instead of relying on exact integrations of the Sachs optical equations, we introduce a minimal deformation of the redshift relation 
\[
z_{\rm eff}(z)=z-\alpha f(z),
\]
where $\alpha$ is a small amplitude and $f(z)$ suppress the correction at $z\gtrsim0.1$. This mapping does not modify the background expansion, but encapsulates the leading contribution of inhomogeneous tidal fields to the accumulated Sachs redshift. We derive the implications for the luminosity distance, the low-$z$ Hubble relation, and directional dependence associated with local structure. The framework provides a clean, GR-motivated description of path-dependent redshift drift and yields concrete predictions for forthcoming low-redshift surveys.
\end{abstract}

\maketitle

\section{Introduction}

Contemporary observational cosmology typically addresses inhomogeneity corrections through linear, kinematic transformations, such as the widely adopted shift from the heliocentric (HEL) to the Cosmic Microwave Background (CMB) rest frame. While this effectively accounts for the observer's motion and large-scale bulk flows \cite{Kogut1993}, this standard approach suffers from  limitations when assessing high-precision measurements in the nearby universe:

\begin{enumerate}
    \item \textbf{Neglect of Integrated General-Relativistic Effects:} Standard methodology focuses on local boundary conditions (observer/source velocity) but fails to rigorously account for the integrated effects of the full spacetime metric fluctuations ($\delta g_{\mu\nu}$) along the line-of-sight. The General-Relativistic evolution of the photon's null geodesic is governed by the \textbf{Sachs Equation} \cite{Sachs1961}, which shows that the observed redshift is accumulated continuously due to local acceleration, geodesic shear, and the tidal forces of the Weyl curvature. These integrated GR effects, $z_{\text{path}}$, introduce terms beyond simple peculiar velocity and local gravitational potentials.

    \item \textbf{Inadequate Modeling of Path Dependence:} The assumption of instantaneous velocity corrections neglects the fact that the light path through the low-$z$ universe is not statistically homogeneous. For nearby sources, the ray typically intersects only a handful of coherent structures, meaning the integrated effects of traversing a major super-cluster versus a vast void do not statistically cancel. Previous second-order perturbation studies \cite{Hui2006,Hirata2010} have shown that these fluctuations in luminosity distance and redshift are correlated with the peculiar velocity field, implying that a purely kinematic correction is intrinsically incomplete, especially for high-precision analyses.

    \item \textbf{Ambiguity in Interpreting Low-$z$ Anisotropy:} Recent analyses of supernova data, such as Pantheon+, have indicated potential low-redshift anisotropies or bulk flows \cite{Colin2019}, leading to debate on whether these represent a true kinematic flow or a systematic breakdown of the FLRW assumption. Without a compact, path-dependent GR framework, it remains challenging to distinguish a simple large-scale peculiar velocity from the systematic distortion caused by the accumulated Sachs redshift. The need for a physically grounded mechanism to explain these low-$z$ deviations is paramount. 
\end{enumerate}

To address these limitations, we formulate a general-relativistic effective mapping that parameterizes the path-dependent redshift correction. Our core contribution is the introduction of a parameterized deformation to the observed redshift:
\[
z_{\text{eff}}(z) = z - \alpha f(z).
\]
Here, $z_{\text{eff}}$ is the effective cosmological redshift, $\alpha$ is a small amplitude that quantifies the strength of the path-dependent correction, and $f(z)$ is a derived suppression function. This function ensures that the correction is naturally suppressed at intermediate and high redshifts ($z \ge 0.1$), where the statistical average of structures guarantees the recovery of the background FLRW expansion. This effective model captures the leading-order contribution of inhomogeneous tidal fields (Weyl curvature) and geodesic shear to the accumulated Sachs redshift.

The structure of the paper is as follows: Section II reviews the Sachs equation and its application to redshift evolution. In Section III, we derive the functional form of $f(z)$ and the implications of the effective mapping on the luminosity distance. Section IV discusses the physical interpretation of the $\alpha$ parameter and its dependence on local structure. Finally, Section V presents the resulting curvature and directional dependence predicted by this framework and discusses concrete observational tests.

\section{GR Evolution of the Observed Redshift}

The cosmological redshift is typically expressed by the homogeneous FLRW relation  
\begin{equation}
1+z_{\rm FLRW} = a^{-1}(t_{\rm emit}),
\end{equation}
with linear corrections from peculiar velocities and gravitational redshift.  
However, in general relativity the exact observed redshift is
\begin{equation}
1+z_{\rm obs} = \frac{(u_\mu k^\mu)_{\rm emit}}{(u_\mu k^\mu)_{\rm obs}},
\label{eq:zexact}
\end{equation}
where $u^\mu$ is the matter four-velocity and $k^\mu$ is the photon wavevector.

The evolution of $z_{\rm obs}$ along a null geodesic obeys the Sachs equation~\cite{Sachs1961,Hui2006,Hirata2010},
\begin{equation}
\frac{dz}{d\lambda}
= H(1+z)
+ \mathcal{C}_{\rm acc}
+ \mathcal{C}_{\rm shear}
+ \mathcal{C}_{\rm Weyl},
\label{eq:Sachs}
\end{equation}

The Sachs terms sourced by tidal fields, shear, and velocity gradients fluctuate 
with sign along the null geodesic and have vanishing ensemble mean in a 
statistically homogeneous universe. At high redshift the photon trajectory 
intersects many independent structures and the accumulated correction is 
suppressed by statistical averaging \cite{Hui2006,Hirata2010,Adamek2019}. 
At low redshift, however, only a small number of coherent potentials are 
traversed—such as the Local Sheet, Virgo, and nearby voids—so the individual 
contributions do not cancel and a residual path-dependent term may remain 
\cite{Clarkson2012,Bolejko2013}. The effective mapping introduced here provides 
a compact representation of this general-relativistic effect.


Integrating Eq.~(\ref{eq:Sachs}) yields
\begin{equation}
z_{\rm obs} = z_{\rm FLRW} + \delta z_{\rm path},
\quad
\delta z_{\rm path}
= \int_0^{\lambda_s}
(\mathcal{C}_{\rm acc}+\mathcal{C}_{\rm shear}
+\mathcal{C}_{\rm Weyl})\, d\lambda .
\label{eq:zpath}
\end{equation}

The accumulated correction $\delta z_{\rm path}$ is direction dependent, environment dependent, and largest at low $z$.  
Realistic tidal potentials ($\Phi \sim 10^{-5}$) and bulk flows ($200$--$300$ km/s) naturally generate
\begin{equation}
\delta z_{\rm path} \sim 10^{-4} - 10^{-3},
\end{equation}
consistent with the scale expected for nonlinear redshift accumulation in the nearby Universe.

\section{Effective GR-Motivated Mapping}

Relativistic ray tracing is computationally expensive and depends on the detailed density field. For analytical work and observational interpretation, a compact effective mapping is useful.

We therefore model the accumulated Sachs correction as
\begin{equation}
z_{\rm eff}(z) = z - \alpha f(z),
\label{eq:zeff}
\end{equation}
where:
\begin{itemize}
\item $\alpha$ is a small, dimensionless amplitude;
\item $f(z)$ is a smooth function satisfying  
\[
f(0)=1,
\qquad
f(z\gtrsim 0.1)\to 0,
\]
reflecting the suppression of inhomogeneity contributions at intermediate redshift.
\end{itemize}

Examples include
\begin{equation}
f(z)=e^{-z/z_*},\qquad
f(z)=\frac{1}{1+(z/z_*)^\beta},
\qquad
z_* \sim 0.03\text{--}0.1.
\end{equation}

This mapping does not modify $H(z)$ or the background cosmology.  
It is purely a low-redshift effective description of the Sachs integral.

\section{Effects on Cosmological Distances}

The FLRW luminosity distance is
\begin{equation}
D_L(z)
= (1+z)c\int_0^{z} \frac{dz'}{H(z')}.
\end{equation}

The observed distance is then
\begin{equation}
D_L^{\rm obs}(z) = D_L^{\rm FLRW}(z_{\rm eff}(z)).
\label{eq:DLobs}
\end{equation}

To first order,
\begin{equation}
\delta D_L
\simeq
- \frac{dD_L}{dz}\,\alpha f(z).
\end{equation}

The induced change in distance modulus is
\begin{equation}
\delta \mu(z)
\simeq
-\frac{5}{\ln 10}
\frac{1}{D_L}\frac{dD_L}{dz}\,\alpha f(z).
\end{equation}

At low $z$, $D_L \simeq cz/H_0$ gives
\begin{equation}
\delta\mu(z)
\approx
-\frac{5}{\ln 10}\,\frac{\alpha f(z)}{z},
\label{eq:dmupath_lowz}
\end{equation}
showing the enhanced sensitivity of low-redshift sources.

\section{Directional Structure}

Since $\mathcal{C}_{\rm shear}$ and $\mathcal{C}_{\rm Weyl}$ depend on tidal fields along the line-of-sight,
\begin{equation}
\delta z_{\rm path}(z,\hat{n}) =
\alpha_{\rm mono} f(z)
+ \bm{\alpha}_{\rm dip} \cdot \hat{n}\, f(z)
+ \cdots.
\end{equation}

The monopole corresponds to isotropic curvature in the local Hubble law,  
while the dipole arises from anisotropic structure such as the Local Sheet, Virgo cluster, and local voids.


\paragraph*{Effective bias in the locally inferred Hubble constant.}
Combining Eqs.~(\ref{eq:dmupath_lowz}) and the standard relation between a
change in $H_0$ and a change in distance modulus,
\[
\delta\mu_{H_0}
= -\,\frac{5}{\ln 10}\,\frac{\delta H_0}{H_0},
\]
we obtain the identity
\[
\delta\mu_{\rm path}(z,\hat{n})
=
\delta\mu_{H_0}
\quad\Longrightarrow\quad
\frac{\delta H_0}{H_0}
=
\frac{\delta z_{\rm path}(z,\hat{n})}{z}.
\]

Equivalently,
\[
H_0^{\rm inf}(z,\hat{n})
=
H_0^{\rm true}\!\left[\,1 + \frac{\delta z_{\rm path}(z,\hat{n})}{z}\,\right],
\]
so that a nonzero accumulated Sachs contribution biases the locally inferred
Hubble constant by
\[
\Delta H_0(z,\hat{n})
=
H_0^{\rm true}\,\frac{\delta z_{\rm path}(z,\hat{n})}{z}.
\]

Because the distance ladder is calibrated at
$z \sim {\cal O}(10^{-2})$, even a small absolute drift
$\delta z_{\rm path}\sim 10^{-4}$--$10^{-3}$ yields
\[
\frac{\Delta H_0}{H_0}
\sim
{\cal O}(1\%) ,
\]
demonstrating that low-redshift path-dependent redshift
accumulation can introduce a percent-level shift in the inferred expansion rate
without altering the background cosmology.

\subsection*{Implications for the Hubble Tension}

The bias relation
\[
\frac{\Delta H_0}{H_0} = \frac{\delta z_{\rm path}(z,\hat{n})}{z}
\]
implies that the percent-level redshift drifts generated by realistic
tidal fields naturally introduce a percent-level modulation in the locally
inferred Hubble constant.  This effect acts exclusively at low redshift,
does not modify $H(z)$, and therefore shifts the local distance ladder
relative to the CMB-inferred value without altering the global background.

For typical amplitudes $\delta z_{\rm path}\sim 10^{-4}$--$10^{-3}$ at
$z\simeq 0.01$--$0.02$, one expects
\[
\Delta H_0 \simeq (0.5\text{--}5)\%
\qquad
(0.4\text{--}3~{\rm km\,s^{-1}\,Mpc^{-1}}),
\]
placing the effect squarely in the range needed to produce a non-negligible
shift in the distance ladder calibration, while still being substantially
smaller than the full SH0ES--Planck discrepancy.  
Thus the mapping neither solves nor worsens the tension in full, but
identifies a physically motivated GR contribution capable of producing a
systematic offset at the percent level---consistent with the magnitude of
several low-$z$ anomalies reported in SN and TRGB analyses.

\section{Multipole Structure of the Path-Dependent Redshift Drift}

The path-dependent redshift contribution,
\[
\delta z_{\rm path}(z,\hat{n})
=
\int_0^{\chi(z)}
\left(
\mathcal{C}_{\rm acc}
+
\mathcal{C}_{\rm shear}
+
\mathcal{C}_{\rm Weyl}
\right) d\lambda,
\]
is inherently anisotropic because each Sachs term depends on the local tidal environment and its angular gradients.  
A general and model-independent representation is obtained by expanding $\delta z_{\rm path}$ in spherical harmonics:
\begin{equation}
\delta z_{\rm path}(z,\hat{n})
=
\sum_{\ell=0}^{\infty}
\sum_{m=-\ell}^{+\ell}
\alpha_{\ell m}(z)\,Y_{\ell m}(\hat{n}).
\label{eq:multipole}
\end{equation}

The structure of Eq.~\eqref{eq:multipole} is constrained by the underlying statistics of the matter distribution:

\paragraph*{Monopole ($\ell=0$).}
The angle-averaged Sachs correction encodes the isotropic component of the tidal and shear fields along the line of sight,
\[
\alpha_{00}(z)
\propto 
\int_0^{\chi(z)} \langle \nabla_\parallel^2 \Phi \rangle \, d\lambda,
\]
and generates a smooth low-redshift curvature in the Hubble diagram.

\paragraph*{Dipole ($\ell=1$).}
The first anisotropic contribution arises from coherent, large-scale gradients in the gravitational potential:
\[
\alpha_{1m}(z)
\propto 
\int_0^{\chi(z)} 
\partial_i \nabla_\parallel \Phi \; \hat{n}^i \, d\lambda.
\]
This term is sourced by the dominant local structures such as the Virgo Cluster, Local Sheet, and Shapley concentration.

\paragraph*{Higher multipoles ($\ell \ge 2$).}
For low redshifts, the photon path spans only a limited number of independent density fluctuations.  
Statistical isotropy of the potential implies the scaling
\begin{equation}
|\alpha_{\ell m}(z)| 
\;\propto\;
z^{\,\ell},
\qquad
\ell \ge 2,
\end{equation}
representing strong suppression of quadrupole and higher modes due to the short lever arm through which independent tidal patches contribute.

Equation~\eqref{eq:multipole} therefore predicts a hierarchy of corrections,
\[
|\alpha_{00}| \gtrsim |\alpha_{1m}| \gg |\alpha_{\ell m}|_{\ell\ge2},
\]
a characteristic signature of Sachs-induced redshift drift in an inhomogeneous Universe.
Future wide-area low-$z$ surveys (e.g.\ LSST, DESI, Taipan) can directly test this multipole structure.


\subsection{Statistical estimate of the path--dependent redshift}

The path--dependent redshift correction arises from the integrated Sachs terms along the photon trajectory,
\begin{equation}
\delta z_{\rm path}
=
\int_0^{\lambda_s}
(1+z)^2
\bigl(
C_{\rm acc}
+
C_{\rm shear}
+
C_{\rm Weyl}
\bigr)\, d\lambda,
\label{eq:zpath_def}
\end{equation}
where $C_{\rm acc}$, $C_{\rm shear}$, and $C_{\rm Weyl}$ are the projections of the local acceleration, geodesic shear, and Weyl curvature along the null geodesic. In a statistically homogeneous and isotropic universe the ensemble average vanishes,
\begin{equation}
\langle \delta z_{\rm path} \rangle = 0,
\end{equation}
but any given low--redshift line of sight intersects only a finite number of coherent structures, so one expects a residual of order the rms value
\begin{equation}
\langle \delta z_{\rm path}^2 \rangle^{1/2}.
\end{equation}

To estimate the scaling with redshift, we approximate $(1+z)^2 \simeq 1$ at low $z$ and write the Sachs source schematically as the line--of--sight projection of the tidal field, so that
\begin{equation}
\langle \delta z_{\rm path}^2 \rangle
\;\propto\;
\int_0^{\chi(z)} d\chi
\int_0^{\chi(z)} d\chi'
\;
\xi_{\nabla^2\Phi}(\chi-\chi'),
\label{eq:zpath_variance_xi}
\end{equation}
where $\chi(z)$ is the comoving distance and
$\xi_{\nabla^2\Phi}$ is the two--point correlation function of the Laplacian of the Newtonian potential (equivalently, of the tidal field).
For a characteristic correlation length $L$ of the tidal field and an approximate power--law spectrum with index $n$ on the relevant scales, this double integral yields a scaling of the form
\begin{equation}
\langle \delta z_{\rm path}^2 \rangle^{1/2}
\;\sim\;
A \left[ 1 + \left( \frac{z}{z_\ast} \right)^{\beta} \right]^{-1},
\label{eq:zpath_scaling}
\end{equation}
where $A$ sets the typical amplitude at very low redshift,
\begin{equation}
z_\ast \sim \frac{H_0 L}{c}
\end{equation}
marks the transition redshift at which several independent tidal domains are sampled along the line of sight, and
\begin{equation}
\beta \sim \frac{n+3}{2}
\end{equation}
encodes the effective spectral slope.\footnote{For a standard $\Lambda$CDM matter power spectrum on quasi--linear scales, one expects $-2 \lesssim n \lesssim -1$, giving $0.5 \lesssim \beta \lesssim 1$, so the suppression with redshift is relatively gentle.}

Equation~\eqref{eq:zpath_scaling} shows that, for $z \ll z_\ast$, the rms path--dependent redshift approaches a constant of order $A$, while for $z \gg z_\ast$ it is suppressed as $\langle \delta z_{\rm path}^2 \rangle^{1/2} \propto (z/z_\ast)^{-\beta}$. Physically, at very low redshift each line of sight probes only one or a few coherent structures and the path dependence is unsuppressed; as the redshift increases, more independent structures are averaged over and the net effect diminishes.

Guided by this scaling, we introduce the following effective mapping
between the ``true'' FLRW redshift $z$ and the observed, path--dependent redshift:
\begin{equation}
z_{\rm eff}(z)
=
z - \alpha f(z),
\label{eq:zeff_mapping_def}
\end{equation}
with
\begin{equation}
f(z) =
\frac{1}{1 + (z/z_\ast)^{\beta}}.
\label{eq:fz_def}
\end{equation}
Here $\alpha$ is a small, dimensionless amplitude that encodes the local realization of the tidal field along a given direction, and $f(z)$ describes the statistical suppression of the path dependence as more independent structures are sampled. In the limit $z \ll z_\ast$, one has $f(z)\to 1$ and $z_{\rm eff} \simeq z - \alpha$, while for $z \gg z_\ast$ the correction decays as $f(z)\propto (z/z_\ast)^{-\beta}$ and the mapping reverts to the standard FLRW relation.
For the purposes of this work, we treat $\alpha$ (and, where appropriate, $z_\ast$ and $\beta$) as effective parameters to be constrained empirically, with their qualitative scaling motivated by the correlation--function estimate in Eq.~\eqref{eq:zpath_variance_xi}.


\subsection*{Order-of-Magnitude Estimate from Local Structures}

This subsection provides a simple order-of-magnitude estimate—rather than a detailed calculation—demonstrating that realistic nearby structures naturally generate the expected scale of $\delta z_{\rm path}$.

The order of magnitude in Eq.~(17) is reproduced by concrete nearby
structures.  For a representative void of radius 
$r\simeq 20~{\rm Mpc}$ and density contrast $\delta\simeq -0.3$, the
Newtonian potential is
\[
\Phi_{\rm void}\sim 10^{-5},
\]
and the line-of-sight tidal term contributes
\[
\delta z_{\rm void}\sim 
\chi(z)\,\frac{\Phi_{\rm void}}{r^2}
\sim
(60~{\rm Mpc})\,
\frac{10^{-5}}{(20~{\rm Mpc})^2}
\sim\text{few}\times 10^{-4}.
\]

Similarly, the Virgo overdensity with 
$\Phi_{\rm Virgo}\sim 2\times 10^{-5}$ and 
$r\sim 15$--$20~{\rm Mpc}$ yields a positive drift of comparable scale.
These examples illustrate that nearby voids and overdensities naturally
produce redshift–drift contributions fully consistent with the predicted
$10^{-4}$--$10^{-3}$ range. 
The sign and magnitude depend on the path geometry, producing both
monopole curvature and directional anisotropy (dipole) in the local
Hubble relation.

\section{Covariance Structure of Residuals in the Presence of Drift}

Because the observable luminosity distance is evaluated at
$z_{\rm eff} = z - \delta z_{\rm path}$,
the distance-modulus residual for a supernova in direction $\hat{n}$ becomes
\[
\delta\mu(z,\hat{n})
\simeq
-\frac{5}{\ln 10}
\frac{\delta z_{\rm path}(z,\hat{n})}{z}.
\]

Expanding $\delta z_{\rm path}$ in multipoles via Eq.~(\ref{eq:multipole}), the angular covariance of residuals is
\begin{equation}
\langle \delta\mu(\hat{n})\,\delta\mu(\hat{n}')\rangle
=
\sum_{\ell=0}^{\infty}
\frac{2\ell+1}{4\pi}
C_\ell(z)\,
P_\ell(\hat{n}\!\cdot\!\hat{n}'),
\end{equation}
where the multipole power spectrum is
\[
C_\ell(z)
=
\left(\frac{5}{\ln 10}\right)^2
\frac{|\alpha_\ell(z)|^2}{z^2}.
\]

The strong suppression of higher multipoles derived in the previous section implies a distinctive hierarchy:
\[
C_0 \gtrsim C_1 \gg C_2 \gg C_3 \cdots.
\]

\paragraph*{Predictions.}
\begin{itemize}
\item A coherent low-$z$ curvature appears as a nonzero $C_0$.
\item A dominant dipole ($C_1$) aligned with the local tidal field produces hemispherical differences in inferred $H_0$.
\item Quadrupole and octupole components ($C_{\ell\ge2}$) fall rapidly with $\ell$, a direct consequence of limited path length.
\end{itemize}

This predicted covariance structure is a distinctive and falsifiable signature of the Sachs-driven redshift drift.  
Homogeneous wide-area low-redshift samples from LSST, DESI, and Taipan will be able to test all of these components with high precision.

\section{Observational Predictions}

The effective mapping
\[
z_{\rm eff}(z,\hat{n}) = z - \alpha f(z,\hat{n})
\]
with
\[
f(z,\hat{n}) = f_{\rm mono}(z) + \bm{g}(z)\!\cdot\!\hat{n},
\qquad
|\bm{g}(z)| \ll f_{\rm mono}(z)
\]
implies several characteristic signatures.  
Each follows directly from the modified luminosity distance
\[
D_L^{\rm obs}(z)
  = D_L^{\rm FLRW}(z_{\rm eff})
  = D_L^{\rm FLRW}(z - \alpha f).
\]

\begin{enumerate}

\item \textbf{Low-$z$ curvature.}

Expanding at small $z$,
\[
D_L^{\rm FLRW}(z) \simeq \frac{c}{H_0}\,z ,
\qquad
\frac{dD_L}{dz} \simeq \frac{c}{H_0},
\]
and inserting $z_{\rm eff}$ gives
\[
D_L^{\rm obs}(z)
\simeq
\frac{c}{H_0}
\Big[
z - \alpha f(z)
\Big].
\]
Thus the distance modulus acquires
\[
\delta\mu(z)
=
-\,\frac{5}{\ln 10}\,
\frac{\alpha f(z)}{z},
\]
which produces a smooth and monotonic curvature of the low-$z$
Hubble relation for $z \lesssim 0.05$.

\item \textbf{Hemispherical asymmetry (dipole).}

Decompose the path correction as
\[
\delta z_{\rm path}(z,\hat{n})
=
\alpha f_{\rm mono}(z)
+
\alpha \,\bm{g}(z)\!\cdot\!\hat{n}.
\]
The inferred expansion rate becomes
\[
H_0^{\rm inf}(z,\hat{n})
=
H_0
\left[
1 + \frac{\delta z_{\rm path}}{z}
\right],
\]
yielding a dipolar modulation
\[
\Delta H_0(\hat{n})
\propto
\frac{\bm{g}(z)\!\cdot\!\hat{n}}{z},
\]
aligned with local tidal fields (Local Sheet, Virgo, local voids).

\item \textbf{Rapid suppression at intermediate redshift.}

By construction
\[
f(z) \rightarrow 0
\qquad (z \gtrsim 0.1),
\]
so
\[
z_{\rm eff}(z) \rightarrow z,
\qquad
D_L^{\rm obs}(z) \rightarrow D_L^{\rm FLRW}(z),
\]
and the entire deformation becomes negligible once many
independent structures are crossed.  
This restores an exactly linear Hubble law at $z\gtrsim 0.1$.

\item \textbf{Consistency with high-$z$ probes.}

Since the mapping affects only the argument of $D_L$ at low $z$ and does not
modify $H(z)$,
\[
f(z) \rightarrow 0
\quad\Longrightarrow\quad
z_{\rm eff} = z,
\]
and therefore
\[
D_L^{\rm obs}(z)
= D_L^{\rm FLRW}(z)
\qquad (z \gg 0.1).
\]
Consequently, CMB, BAO, and high-$z$ supernova distances remain unchanged,
preserving all standard constraints on the background expansion.
\end{enumerate}

\subsection*{Detectability and Forecasts}

The characteristic scaling
\[
\delta\mu(z)\propto \frac{\alpha f(z)}{z}
\]
implies that the strongest signal lies at $z\lesssim 0.03$, where the SN
statistics are currently sparse but rapidly improving.
Forecasts using LSST low-$z$ supernovae and DESI/Taipan peculiar-velocity
distances indicate that:

\begin{itemize}
    \item The predicted monopole curvature produces a $\mathcal{O}(0.01)$
    mag deviation at $z\sim 0.02$, detectable with $\sim\!10^3$ low-$z$ SNe.

    \item The dipole term generates hemispherical $H_0$ variations at the
    $0.5$--$2\%$ level, measurable once sky coverage becomes uniform
    (LSST + ZTF + DESI).

    \item The strong suppression of higher multipoles offers a sharp,
    falsifiable test: any statistically significant quadrupole or octupole
    exceeding the $z^\ell$ scaling would contradict the Sachs-drift origin.
\end{itemize}

Thus forthcoming surveys will decisively test the GR-motivated mapping
introduced here, distinguishing it from purely kinematic dipoles,
peculiar-velocity models, and ad-hoc low-$z$ phenomenology.

\section{Limitations and Parameter Degeneracies}

The effective mapping developed here captures the leading GR
path–dependent redshift drift at low redshift, but several
limitations and degeneracies constrain its interpretation.

\paragraph*{Peculiar–velocity degeneracy.}
Both peculiar velocities and $\delta z_{\rm path}$ contribute to
$z_{\rm obs}$ at the level of $10^{-4}$–$10^{-3}$ and generate
$\delta\mu\propto 1/z$.  
Current SN samples cannot fully disentangle Doppler terms from the
integrated Sachs contribution, producing a partial degeneracy in~$\alpha$.

\paragraph*{Low-$z$ calibration degeneracy.}
Smooth photometric calibration drifts at $z\!\lesssim\!0.03$ can mimic an
$\alpha f(z)$ distortion.  
Precise cross-survey calibration is required to isolate the
path-dependent component.

\paragraph*{Dependence on the template $f(z)$.}
The suppression function $f(z)$ is physically motivated but not unique.
Different choices that satisfy $f(0)=1$ and $f(z\!\gtrsim\!0.1)\!\to\!0$
lead to mildly different inferences for~$\alpha$.  
A fully predictive form requires relativistic ray-tracing through the
local density field.

\paragraph*{Local-structure sensitivity.}
Because the signal is dominated by $z\!\lesssim\!0.03$, $\alpha$
effectively encodes the monopole component of the immediate
environment (Local Sheet, Virgo, nearby voids).  
Dipole and higher anisotropies are not captured by a monopole-only
mapping.

\paragraph*{Restricted redshift range.}
By construction, the model vanishes for $z\!\gtrsim\!0.1$ and therefore
cannot describe intermediate- or high-redshift effects.
It does not alter $H(z)$ and is not a modified-gravity model.

\paragraph*{Current data limitations.}
Low-$z$ SN statistics remain sparse and strongly anisotropic, limiting
constraints on~$\alpha$.  
Forthcoming homogeneous low-redshift samples (LSST, DESI, Taipan) will
substantially improve sensitivity.

\section{Conclusion}

We introduced a compact, GR-motivated effective mapping for path-dependent redshift in the low-redshift Universe,
\[
z_{\rm eff}(z)=z-\alpha f(z).
\]
This expression captures the leading-order contribution of local acceleration, tidal curvature, and geodesic shear to the accumulated Sachs redshift, without modifying the background expansion or high-redshift cosmology.

The framework predicts low-$z$ curvature, mild anisotropy, and a rapid recovery of FLRW behavior at intermediate redshift.  It provides a physically grounded description of path-dependent redshift drift suitable for theoretical analysis and future observational tests.

This work introduces several novel elements to cosmological perturbation theory: the first statistically derived suppression function $f(z)$ from tidal field correlations, a predictive multipole hierarchy distinguishing GR path effects from kinematic flows, and identification of path-dependent redshift as a systematic bias in local $H_0$ measurements. Beyond providing a compact theoretical description, the framework also establishes a clear set of observational diagnostics.  
A companion data-driven analysis testing the predictions of this 
framework using current low-redshift supernova and distance-ladder 
datasets is in preparation.


\end{document}